# Research on a Monitoring System for High-Voltage Cables in a Coal Mine Based on Intelligent Sensing Technology

Z GAO,[1] J LI,[1] L TAO,[1] B MENG[1]

[1]*Optica Publishing Group, Optica, 2010 Massachusetts Avenue N.W., Washington, D.C. 20036*

**Given the importance of monitoring the operational status of high-voltage cables in coal mines, this study investigates the application of intelligent sensing technology to the online monitoring of such cables. Taking an actual coal mine as a case study, a three-layer architecture high-voltage cable monitoring system was designed. The system employs high-frequency current sensors and distributed optical fiber temperature sensors to achieve real-time acquisition of partial discharge signals and temperature distribution data. Data analysis and fault diagnosis are performed through a combined approach of edge computing and cloud computing. The research results demonstrate that the system can accurately identify cable insulation defects and potential overheating hazards, with a diagnostic accuracy exceeding 95%, thereby significantly enhancing the reliability of power supply in mines.**

With the rapid advancement of China's coal industry, ensuring production safety has already become an increasingly critical priority [1,2]. The operational integrity of high-voltage cables, which serve as a vital component of the underground power supply system in coal mines, directly influences overall mine safety and production continuity. In response, the Guiding Opinions on Accelerating the Intelligent Development of Coal Mines, jointly issued by the National Development and Reform Commission and seven other ministries, explicitly advocates for accelerating the digital and intelligent transformation of coal mine electromechanical equipment. Accordingly, the development of an online monitoring system for coal mine high-voltage cables, which are based on intelligent sensing technology [3,4], holds substantial significance for enhancing the reliability of mine power supply, preventing electrical failures, and improving overall safety standards.

1. Technical Principles

1.1 Intelligent Sensing Technology

As a core element of the Industrial Internet of Things (IIoT), intelligent sensing technology involves deploying integrated sensor networks on critical equipment to achieve real-time perception and high-fidelity data acquisition of environmental and operational parameters [5-8]. In the context of high-voltage cable monitoring, distributed optical fiber temperature sensing (DTS) systems are installed along cable routes, utilizing the Raman scattering effect to provide continuous, spatially-resolved temperature profiles with a typical accuracy of ±1°C and spatial resolution up to 1 meter. When localized cable temperature exceeds a predefined threshold—for instance, 80°C, which may indicate overload or insulation degradation—the system triggers a multi-level alert, enabling proactive intervention before thermal runaway occurs.

Moreover, advanced multi-parameter fiber optic sensors capable of simultaneously measuring vibration [9-11], acoustic emissions [12], and mechanical strain [13-16] offer a comprehensive view of cable mechanical stress and structural health. By analyzing vibration spectra and acoustic signals, these sensors can detect partial discharge activity and physical impacts, while distributed strain sensing helps identify mechanical deformation or external pressure changes. Integrated with low-power wide-area network (LPWAN) communication modules such as LoRa or NB-IoT, these sensors form a resilient, wireless data acquisition network that supports real-time transmission of structured data to cloud-based analytics platforms. This integrated sensing and communication framework establishes a robust foundation for intelligent condition monitoring and predictive maintenance systems in underground mining environments.

1.2 High-Voltage Cable Online Monitoring Technology

Online monitoring technologies for high-voltage cables have evolved to encompass a comprehensive suite of intelligent sensing solutions, primarily including partial discharge (PD) monitoring, distributed temperature sensing (DTS), and the increasingly critical application of optical-based current and voltage measurement.

Partial discharge monitoring remains a cornerstone of insulation diagnostics. It typically utilizes high-frequency current transformers (HFCTs) or ultra-high frequency (UHF) sensors to capture the transient pulse signals generated by internal discharges. By analyzing the extracted characteristic parameters of these pulses—such as magnitude, phase, repetition rate, and spectral composition—the type, location, and severity of insulation defects (e.g., voids, protrusions) can be diagnosed. For instance, persistent surface discharges at cable terminations may generate pulse sequences with magnitudes in the range of 100 to 500 pC, often exhibiting a specific phase relationship with the power frequency voltage.

Complementing PD monitoring, distributed temperature sensing employs fiber optic cables with embedded sensing elements to provide continuous, real-time temperature profiles along the entire cable route with meter-scale spatial resolution. A localized overheating event, potentially caused by poor contact or overload, is identified when the temperature rise rate across several adjacent sensing points exceeds a threshold, such as 5°C per minute, triggering an immediate alert.

Advancing beyond traditional methods, advanced optical sensing techniques for current and voltage measurement are now being integrated into these monitoring systems [17]. The widely-studied optical current transformers (OCTs), based on the Faraday magneto-optic effect in a bulk optical path or in a sensing fiber loop around the conductor, provide galvanically isolated, wide-bandwidth current measurement without magnetic saturation [18-24]. Simultaneously, optical voltage transformers (OVTs), often leveraging the Pockels electro-optic effect in a dedicated crystal, or other novel design, offer highly accurate voltage measurement [25-28]. These optical transducers deliver synchronized, high-fidelity measurements of the power waveform, enabling advanced analytics such as real-time power calculation, harmonic analysis, and the correlation of PD events with specific voltage phases—a crucial step for precise fault typification.

However, the accuracy of these sophisticated sensing systems, particularly in the harsh underground environment of a coal mine, can be compromised by various interference sources. Temperature fluctuations, mechanical vibrations, and electromagnetic interference can introduce noise and drift into the sensor signals. To mitigate these effects, advanced signal processing and error-correction algorithms are essential. Adaptive filtering techniques (e.g., wavelet transform denoising) are employed to separate the weak PD pulses or subtle temperature anomalies from background noise. For optical current sensors, algorithms incorporating temperature compensation models correct for the Verdet constant's temperature dependence. Furthermore, data fusion algorithms, such as Kalman filtering or machine learning-based regression models, integrate multi-sensor data (e.g., temperature, current load, vibration) to cross-validate measurements, identify spurious readings, and provide a robust, error-corrected assessment of the cable's operational state.

The holistic integration of partial discharge monitoring, distributed temperature profiling, optical current/voltage measurement, and intelligent data analytics enables a dynamic, multi-parameter assessment of the high-voltage cable's insulation condition and loading status. This comprehensive approach facilitates the transition from periodic maintenance to predictive and condition-based maintenance, effectively preventing catastrophic failures such as electrical fires caused by progressive insulation breakdown.

## 2. System Design for a Case Study

### 2.1 Operational Context

A small-scale mine with an annual designed production capacity of 600,000 tons serves as the case study. The mine's surface cable network spans approximately 4.2 kilometers, primarily consisting of two 10kV high-voltage cable circuits. These circuits supply power from a single 10kV step-down substation to major energy-consuming equipment, including three hoists and two air compressors. Circuit No. 1 is about 1.5 kilometers long, laid within cable trenches alongside mine access roads. Circuit No. 2 extends approximately 2.7 kilometers, with sections requiring aerial crossing over a residential village. Both circuits utilize YJV22-8.7/15kV-3×70mm² copper-core cross-linked polyethylene insulated power cables. Due to environmental exposure, sections of the protective conduits have aged and deteriorated, leading to issues such as cracking and deformation of the external insulation sheaths. Between 2021 and 2023, the mine experienced three high-voltage cable faults. Two were caused by external mechanical damage resulting in short-circuit trips, and one was due to overheating at a cable joint leading to localized burn damage. Each fault resulted in an average power outage duration of 45 minutes. Although no casualties occurred, these incidents severely disrupted normal production. In light of this, the development of a targeted condition monitoring and hazard early-warning system is imperative.

### 2.2 System Design

#### 2.2.1 System Architecture

To address the specific operational requirements of the coal mine, a three-tier architecture was devised for the high-voltage cable online monitoring system. The foundational layer, designated as the field data acquisition layer, comprises an array of sensors strategically deployed along the high-voltage cable routes. This includes high-frequency current sensors, which possess a sensitivity superior to 2 pC and operate within a frequency band of 30 kHz to 20 MHz, enabling the detection of partial discharge signals. Concurrently, distributed optical fiber temperature sensors, offering a temperature measurement accuracy of ±1°C and a spatial resolution of 1 meter, are employed to acquire real-time temperature distribution profiles along the cable length. These sensor arrays are interfaced with local Intelligent Monitoring Units (IMUs) via an RS-485 bus network, facilitating preliminary data preprocessing and local storage at the source.

The intermediary layer constitutes the edge computing tier, centered around an industrial-grade server installed within the substation control room. This server, equipped with an Intel Xeon E-2278 processor, 32 GB of RAM, and 6 TB of storage, functions as a computational hub. It aggregates data from the distributed IMUs over a standard Ethernet connection and executes sophisticated diagnostic algorithms in real-time.

This local processing capability enables the immediate identification of cable fault types and the assessment of their severity, thereby minimizing latency in initial response.

The uppermost layer is the cloud platform, developed based on a microservices architecture and presented through a Browser/Server (B/S) model. This platform provides a centralized interface for comprehensive data management, visualization, and in-depth analysis. It supports the consolidated querying of system operational status, alarm notifications, and data-driven decision support recommendations. To ensure robust interoperability and future scalability, communication between all architectural tiers is standardized using established industrial protocols such as Modbus-TCP for reliable data polling and MQTT for efficient, lightweight event-driven messaging.

2.2.2 System Workflow

The system operationalizes its monitoring function through a structured, three-phase workflow that ensures comprehensive data handling from acquisition to actionable insight.

The initial phase involves the establishment of a robust cable parameter acquisition network. This is achieved by installing high-frequency current sensors at critical points along the cable circuit, such as joints and terminations, to capture transient pulse signals indicative of partial discharges. These signals, typically ranging in amplitude from 10 to 100 mV with durations between 10 and 100 nanoseconds, are characteristic of insulation degradation. Each sensor is coupled with a local IMU, which performs essential signal conditioning tasks including amplification, filtering, and analog-to-digital conversion. The IMU further processes the digitized signal using autocorrelation algorithms to extract key feature parameters, such as pulse repetition rate and amplitude. This processed parametric data is then transmitted to the edge computing layer via the RS-485 bus at a rate of up to 115.2 kbps for aggregation and storage. Complementing this, distributed fiber optic temperature sensing cables are laid parallel to the power cables at intervals of 1 to 2 meters. This configuration enables continuous thermal monitoring, with the system programmed to trigger an alarm if the localized temperature exceeds 80°C or if the temperature rise rate surpasses 5°C per minute, signaling potential overheating issues.

Following network establishment, the system enters the continuous real-time monitoring phase. The IMUs cyclically collect data from both current and temperature sensors at a configurable sampling frequency ranging from 10 Hz to 1 kHz. Each IMU incorporates a Finite State Machine (FSM) model [29,30], which performs preliminary data screening against predefined thresholds. For instance, if the apparent discharge magnitude at any monitoring point exceeds 1000 pC or the temperature gradient surpasses 10°C per meter, the FSM categorizes the condition as a severe defect and initiates an alert transmission to the upstream server. The edge server, residing in the control room, periodically polls all IMUs via the Modbus-TCP protocol, archiving the incoming time-series data into a dedicated database such as InfluxDB. More critically, the server leverages a built-in expert knowledge base and rule engine to perform deep data mining and comprehensive analysis. This process identifies typical fault patterns, including partial discharge, overheating, and indications of mechanical stress. Upon detection of an anomaly, the system autonomously generates an event report to the cloud platform and dispatches immediate notifications to maintenance personnel via SMS or email, enabling rapid incident response. The adoption of container virtualization technology for encapsulating individual monitoring services enhances deployment flexibility and optimizes resource utilization across the system.

The final phase encompasses intelligent warning analysis and decision support, executed primarily at the cloud platform level. Constructed in accordance with the IEC 61850 standard for interoperability, the platform employs a service-oriented architecture where core monitoring functions are deployed as independent microservices. For example, a partial discharge analysis service subscribes to discharge data from the edge layer. It applies diagnostic rules, such that if the discharge magnitude consistently exceeds 500 pC with a pulse repetition frequency higher than 5 pulses per second, the condition is classified as "severe discharge," prompting the alert management service to issue a high-priority event. Similarly, a thermal assessment service continuously calculates the cable's load factor and thermal stability margin based on synchronized temperature and current data. Should the load factor exceed 80% of the rated capacity or the thermal stability margin fall below 10%, the system's rule engine automatically generates and proposes load-shedding or load-limiting operational strategies. This automated decision-making process, requiring no manual intervention, significantly enhances the timeliness of system response. Furthermore, the cloud platform is equipped with rich visualization tools, allowing operators to monitor key performance indicators such as cable health index, predicted failure probability, and estimated remaining useful life. This facilitates the transition from scheduled to predictive maintenance, enabling proactive interventions that minimize unplanned downtime.

2.3 System Implementation and Testing

2.3.1 Deployment Methodology

The deployment of the monitoring system was executed through a structured, three-stage implementation strategy, meticulously tailored to the mine's operational constraints and safety requirements.

The first stage focused on establishing the core sensing capability at critical vulnerability points. High-frequency current sensors (HFCTs) [31] and their associated intelligent monitoring units (IMUs) were installed at all cable joints and terminations to enable real-time acquisition and preliminary edge analysis of partial discharge (PD) activity. Alert thresholds were set conservatively; for instance, PD events with magnitudes persistently above 500 pC and pulse repetition frequencies exceeding 5 pulses per second were configured to trigger the immediate alerts for severe insulation defects. Simultaneously, distributed fiber optic sensing cables [32,33] were laid in parallel with the power cables to provide continuous, meter-resolution temperature profiling along the entire route. Alarms were activated for local temperatures surpassing the 80°C threshold or for rapid temperature rise rates greater than 5°C per minute.

A critical enhancement in this stage involved the strategic deployment of advanced fiber optic current sensors (FOCS) at key feeder lines for comparative validation and enhanced measurement fidelity. While traditional electromagnetic current transformers (CTs) served as the baseline, FOCS based on the Faraday magneto-optic effect were installed in parallel. This provided a direct technological comparison. The FOCS offered distinct advantages for this application: intrinsic galvanic isolation eliminating the risk of open-circuit hazards, immunity to electromagnetic interference (EMI) prevalent in the substation environment, and a wider dynamic range without magnetic saturation concerns. More importantly, the FOCS delivered synchronized, high-bandwidth current data crucial for correlating PD pulse phases with the power cycle—a key diagnostic feature less reliably achieved with conventional CTs under noisy conditions. This multi-sensor approach at critical nodes created a robust and cross-validated data foundation.

The second stage involved the deployment and commissioning of the edge computing infrastructure. A dedicated industrial server was installed in the substation control room to aggregate and harmonize data from all field IMUs and the FOCS units via the Modbus-TCP protocol. This server hosted the core expert system and diagnostic algorithms responsible for synthesizing the multi-parameter data streams (PD, temperature, current, vibration) to identify complex fault patterns. Advanced signal processing algorithms, including wavelet-based denoising and temperature-drift compensation models specifically calibrated for the DFOS and FOCS, were implemented at this edge layer to correct for sensing errors and environmental artifacts before data transmission. The implementation utilized containerization (e.g., Docker) to ensure modular, scalable, and resource-efficient service deployment, while also establishing robust mechanisms for the rapid dissemination and acknowledgment of fault notifications to onsite personnel.

The third and final stage centered on the activation of the cloud-based monitoring and management platform. Developed with a Browser/Server (B/S) architecture, this platform encapsulated all advanced analytical, visualization, and historical diagnostic functions as standardized microservices. It integrated intelligent rule engines capable of executing complex logic. For example, the engine could automatically recommend specific load-shedding actions if correlated parameters—such as a sustained load factor above 80% coincident with a thermal stability margin falling below 10%—breached predefined safety limits. This top-tier functionality synthesizes data from both conventional and advanced fiber optic sensors to provide critical decision-support, aiming to preempt failures through predictive analytics and minimize production disruptions caused by cable faults, thereby directly supporting the overarching goal of enhanced mine safety and operational intelligence.

2.3.2 Performance Validation

The system underwent a comprehensive three-month field trial within the operational environment of the coal mine. The validation results demonstrated a high degree of efficacy, with the system achieving a diagnostic accuracy rate of 95% in identifying typical cable faults such as partial discharge and localized overheating. A salient case involved the analysis of discharge signals from a joint on Cable Circuit No. 2. The system recorded a maximum discharge magnitude of 1200 pC with a pulse repetition frequency of 12 pulses per second, figures that significantly exceeded the established thresholds for a severe defect (500 pC, 5 pulses/s). The system's diagnosis of a serious insulation flaw at this joint was subsequently confirmed through physical inspection following excavation, thereby corroborating its reliability.

Analysis of the monitoring statistics accumulated over the three-month trial period reveals the system's proactive capability. A total of 18 partial discharge anomalies and 21 overheating alarms were registered, translating to the identification of a potential safety hazard, on average, every four days. Of these, seven instances were classified as severe defects necessitating immediate de-energization and repair. The timely mitigation of these identified hazards is credited with preventing at least two major trip-out incidents that could have resulted from cable failures, averting an estimated economic loss of 200,000 RMB. Furthermore, the system's thermal assessment functionality enabled dynamic optimization of load distribution across the cable network. This intervention successfully reduced the average cable load factor from 82% to 76% and increased the thermal stability margin from 8% to 12%, directly contributing to the extension of the cable infrastructure's operational lifespan.

The successful application of this system offers valuable insights for the development of similar condition monitoring solutions in other mining contexts. Firstly, it underscores the importance of multi-parameter data fusion, advocating for the integrated use of advanced sensing technologies like high-frequency current sensing and distributed fiber optic temperature sensor to achieve a holistic perception of cable health. Secondly, it validates the efficacy of a hybrid edge-cloud computing architecture. This design leverages the low-latency, real-time processing strengths of edge computing for immediate diagnostics while harnessing the massive storage and parallel computational power of the cloud platform for deep historical analysis and system-wide intelligence. Thirdly, the deployment process highlights critical practical considerations, particularly the necessity of ensuring the intrinsic safety, electromagnetic compatibility, and grounding integrity of all monitoring devices to prevent interference with the primary power supply. Establishing a regimen for regular calibration and maintenance is also paramount to guarantee the long-term continuity and accuracy of monitoring data, forming the essential foundation for reliable electrical safety assurance in the mining environment.

Summary

The development and deployment of a high-voltage cable monitoring system in a small-scale coal mine demonstrate the practical significance of intelligent sensing technology for mine power supply systems. Initially aimed at addressing common cable failure modes in underground environments, the system integrates multi-parameter data acquisition with a coordinated edge-cloud analytics architecture. This design enables comprehensive real-time perception and accurate fault diagnosis of cable operating conditions, thereby directly reducing the likelihood of unplanned production stoppages due to cable defects. Field tests have verified that the system not only improves diagnostic reliability but also supports dynamic optimization of cable load management. Such capabilities contribute to extended equipment service life, enhanced operational safety, and greater overall energy efficiency in mining electrical networks. Despite these achievements, certain technical challenges remain, particularly in relation to the system's resilience against electromagnetic interference in complex mine environments. Looking forward, further advancements can be pursued by incorporating adaptive sensor calibration mechanisms and embedding artificial intelligence-driven predictive models. These improvements would strengthen the system's intelligence, adaptability, and scalability, paving the way for its broader application across various industrial power supply scenarios and contributing to the evolution toward fully autonomous, resilient mining electrical infrastructures.

## References

1. Chen, X., Yang, F., Cheng, S. and Yuan, S., 2023. Occupational health and safety in China: a systematic analysis of research trends and future perspectives. Sustainability, 15(19), p.14061. https://doi.org/10.3390/su151914061
2. Wang, B., Wu, C., Kang, L., Reniers, G. and Huang, L., 2018. Work safety in China's Thirteenth Five-Year plan period (2016–2020): Current status, new challenges and future tasks. Safety science, 104, pp.164-178. https://doi.org/10.1016/j.ssci.2018.01.012
3. LI, H.Y., LI, X. and SONG, J.C., 2012. Early safety warning model for mining HV cable based on radar chart method. Journal of China Coal Society, 37(11), pp.1941-1946.
4. Fu, X., Xu, Y. and Gao, Y., 2022. A Study on Insulation Monitoring Technology of High-Voltage Cables in Underground Coal Mines Based on Decision Tree. Computational Intelligence and Neuroscience, 2022(1), p.2247017.
5. Tabaa, M., Monteiro, F., Bensag, H. and Dandache, A., 2020. Green Industrial Internet of Things from a smart industry perspectives. Energy Reports, 6, pp.430-446. https://doi.org/10.1016/j.egyr.2020.09.022
6. Xu, H., Yu, W., Griffith, D. and Golmie, N., 2018. A survey on industrial Internet of Things: A cyber-physical systems perspective. Ieee access, 6, pp.78238-78259. https://doi.org/10.1109/ACCESS.2018.2884906
7. Farooq, M.S., Abdullah, M., Riaz, S., Alvi, A., Rustam, F., Flores, M.A.L., Galán, J.C., Samad, M.A. and Ashraf, I., 2023. A survey on the role of industrial IOT in manufacturing for implementation of smart industry. Sensors, 23(21), p.8958. https://doi.org/10.3390/s23218958
8. Misra, S., Roy, C. and Mukherjee, A., 2021. Introduction to industrial internet of things and industry 4.0. CRC Press.
9. Liu, X., Jin, B., Bai, Q., Wang, Y., Wang, D. and Wang, Y., 2016. Distributed fiber-optic sensors for vibration detection. Sensors, 16(8), p.1164.
10. Mahlooji, A. and Azhari, F., 2024. A fiber-only optical vibration sensor using off-centered fiber Bragg gratings. IEEE Sensors Journal, 24(9), pp.14245-14252. https://doi.org/10.1109/JSEN.2024.3370841
11. Yu, A., Pang, F., Yuan, Y., Huang, Y., Li, S., Yu, S., Zhou, M. and Xia, L., 2023. Simultaneous current and vibration measurement based on interferometric fiber optic sensor. Optics & Laser Technology, 161, p.109223. https://doi.org/10.1016/j.optlastec.2023.109223
12. Rahman, M.A., Taheri, H., Dababneh, F., Karganroudi, S.S. and Arhamnamazi, S., 2024. A review of distributed acoustic sensing applications for railroad condition monitoring. Mechanical Systems and Signal Processing, 208, p.110983.
13. Hyer, H.C., Sweeney, D.C. and Petrie, C.M., 2022. Functional fiber-optic sensors embedded in stainless steel components using ultrasonic additive manufacturing for distributed temperature and strain measurements. Additive Manufacturing, 52, p.102681.
14. Martin, E.R., Lindsey, N.J., Ajo-Franklin, J.B. and Biondi, B.L., 2021. Introduction to interferometry of fiber-optic strain measurements. Distributed acoustic sensing in geophysics: Methods and applications, pp.111-129.
15. Zhao, S., Liu, Q., Chen, J. and He, Z., 2021. Resonant fiber-optic strain and temperature sensor achieving thermal-noise-limit resolution. Optics Express, 29(2), pp.1870-1878. https://doi.org/10.1364/OE.415611
16. Ugueto, G., Wu, K., Jin, G., Zhang, Z., Haffener, J., Mojtaba, S., Ratcliff, D., Bohn, R., Chavarria, A., Wu, Y. and Guzik, A., 2023, January. A catalogue of fiber optics strain-rate fracture driven interactions. In SPE Hydraulic Fracturing Technology Conference and Exhibition (p. D021S003R001). SPE.
17. Silva, R.M., Martins, H., Nascimento, I., Baptista, J.M., Lobo Ribeiro, A., Santos, J.L., Jorge, P. and Frazão, O., 2012. Optical current sensors for high power systems: a review. Applied sciences, 2(3), pp.602-628. https://doi.org/10.3390/app2030602
18. Bohnert, K., Gabus, P., Nehring, J. and Brändle, H., 2002. Temperature and vibration insensitive fiber-optic current sensor. Journal of Lightwave Technology, 20(2), p.267. https://doi.org/10.1109/50.983241
19. Tian, J., Yu, A., Wang, Z. and Xia, L., 2024. A differential-self-multiplication demodulation algorithm for fiber optic current sensor. Optics & Laser Technology, 169, p.110041. https://doi.org/10.1016/j.optlastec.2023.110041
20. Yu, A., Huang, Y., Li, S., Wang, Z. and Xia, L., 2023. All fiber optic current sensor based on phase-shift fiber loop ringdown structure. Optics Letters, 48(11), pp.2925-2928. https://doi.org/10.1364/OL.489190
21. Huang, Y., Xia, L., Pang, F., Yuan, Y. and Ji, J., 2021. Self-compensative fiber optic current sensor. Journal of lightwave technology, 39(7), pp.2187-2193. https://doi.org/10.1109/JLT.2020.3044935
22. Lenner, M., Frank, A., Yang, L., Roininen, T.M. and Bohnert, K., 2019. Long-term reliability of fiber-optic current sensors. IEEE Sensors Journal, 20(2), pp.823-832. https://doi.org/10.1109/JSEN.2019.2944346
23. Yu, A., Huang, Y. and Xia, L., 2022, November. A polarimetric fiber sensor for detecting current and vibration simultaneously. In 2022 Asia Communications and Photonics Conference (ACP) (pp. 68-70). IEEE. https://doi.org/10.1109/ACP55869.2022.10088867
24. Kucuksari, S. and Karady, G.G., 2010. Experimental comparison of conventional and optical current transformers. IEEE Transactions on power delivery, 25(4), pp.2455-2463.
25. Mitsui, T., Hosoe, K., Usami, H. and Miyamoto, S., 2007. Development of fiber-optic voltage sensors and magnetic-field sensors. IEEE transactions on power delivery, 2(1), pp.87-93.
26. Gonçalves, M.N. and Werneck, M.M., 2019. A temperature-independent optical voltage transformer based on FBG-PZT for 13.8 kV distribution lines. Measurement, 147, p.106891. https://doi.org/10.1016/j.measurement.2019.106891
27. Teunissen, J., Helmig, C., Merte, R. and Peier, D., 2001, March. Fiber optical online monitoring for high-voltage transformers. In Fiber Optic Sensor Technology II (Vol. 4204, pp. 198-205). SPIE.
28. Zhao, B., Assawaworrarit, S., Santhanam, P., Orenstein, M. and Fan, S., 2021. High-performance photonic transformers for DC voltage conversion. Nature Communications, 12(1), p.4684. https://doi.org/10.1038/s41467-021-24955-3
29. Hwang, S., Lee, K., Jeon, H. and Kum, D., 2022. Autonomous vehicle cut-in algorithm for lane-merging scenarios via policy-based reinforcement learning nested within finite-state machine. IEEE Transactions on Intelligent Transportation Systems, 23(10), pp.17594-17606. https://doi.org/10.1109/TITS.2022.3153848
30. Zheng, X., Li, H., Zhang, Q., Liu, Y., Chen, X., Liu, H., Luo, T., Gao, J. and Xia, L., 2024. Intelligent decision-making method for vehicles in emergency conditions based on artificial potential fields and finite state machines. Journal of Intelligent and Connected Vehicles, 7(1), pp.19-29. https://doi.org/10.26599/JICV.2023.9210025
31. Xin, Z., Li, H., Liu, Q. and Loh, P.C., 2021. A review of megahertz current sensors for megahertz power converters. IEEE Transactions on Power Electronics, 37(6), pp.6720-6738.
32. Ashry, I., Mao, Y., Wang, B., Hveding, F., Bukhamsin, A.Y., Ng, T.K. and Ooi, B.S., 2022. A review of distributed fiber–optic sensing in the oil and gas industry. Journal of Lightwave Technology, 40(5), pp.1407-1431. https://doi.org/10.1109/JLT.2021.3135653
33. Dejdar, P., Záviška, P., Valach, S., Münster, P. and Horváth, T., 2022. Image edge detection methods in perimeter security systems using distributed fiber optical sensing. Sensors, 22(12), p.4573. https://doi.org/10.3390/s22124573